\documentclass[12pt,a4paper]{article}
\pdfoutput=1 %
\usepackage[utf8]{inputenc}
\usepackage{fullpage}
\usepackage{cite}
\usepackage{subcaption}
\usepackage[english]{babel}
\usepackage{amsmath}
\usepackage{physics}
\usepackage{amsthm}
\usepackage{amsfonts}
\usepackage{amssymb}
\usepackage{bbold}
\usepackage{mathrsfs}
\usepackage{mathbbol}
\usepackage[toc,page]{appendix}
\usepackage{subcaption}
\usepackage{graphicx}
\usepackage{bbding}
\usepackage{esvect}
\usepackage{commath}
\usepackage{enumitem}
\usepackage[affil-it]{authblk}
\usepackage{siunitx}
\usepackage{braket}
\usepackage{phaistos}
\usepackage{multicol}
\usepackage{overpic}
\usepackage{comment}
\usepackage[colorlinks=true
  ,urlcolor=blue
  ,anchorcolor=blue
  ,citecolor=blue
  ,filecolor=blue
  ,linkcolor=blue
  ,menucolor=blue
  ,pagecolor=blue
  ,linktocpage=true
  ,pdfproducer=medialab
  ,pdfa=true
]{hyperref}
\usepackage{xcolor}

\usepackage{jheppub}
\usepackage{qcircuit}
\usepackage{float}
\usepackage{tikz}
\usetikzlibrary{backgrounds, positioning,calc}

\newcommand{\Hil}{\mathcal{H}}
 
\title{Towards Entropic Constraints on Quantum Speedups}
\author[a]{Jason Pollack}
\author[b]{and Dylan VanAllen}

\affiliation[a]{Department of Electrical Engineering and Computer Science, Syracuse University, NY 13210, USA}
\affiliation[b]{Department of Physics, Syracuse University, NY 13210, USA}

\emailAdd{japollac@syr.edu}
\emailAdd{djvanall@syr.edu}

\abstract{Some quantum algorithms have ``quantum speedups'': improved time complexity as compared with the best-known classical algorithms for solving the same tasks. Can we understand what fuels these speedups from an entropic perspective? Information theory gives us a multitude of metrics we might choose from to measure how fundamentally `quantum' is the behavior of a quantum computer running an algorithm. The entanglement entropies for subsystems of a quantum state can be analyzed for subsystems of qubits in a quantum computer throughout the running of an algorithm. Here, a framework for making this entropic analysis is constructed, and performed on a selection of quantum circuits implementing known fast quantum algorithms and subroutines: Grover search, the quantum Fourier transform, and phase estimation. Our results are largely unsatisfactory: known entropy inequalities do not suffice to identify the presence or absence of quantum speedups. Although we know our algorithms must have quantum ``magic,'' the Ingleton inequality, which holds for all entropies of subsystems of stabilizer states, is not violated in any of our examples. In some cases, however, monogamy of mutual information, which is obeyed for product states but violated for highly entangled bipartite states such as the $GHZ$ states, fails at some point in the course of our quantum circuits.}
 
\begin{document}
\maketitle

\section{Introduction}
It is well known that the Shannon entropies of probability distributions confined to subsystems are bounded by information theory \cite{Shannon:1948dpw,681320,4557201}. The resulting constraints on the sharing of information between these subsystems define the classical entropy cone. In \emph{quantum} systems, due to entanglement, we know that the quantum entropy cone is different from the classical one \cite{Araki:1970ba,Lieb_1973}: for example, entropy can decrease as a system is made larger, as a result of entanglement between portions of the system. 

In this work, we initiate a study of the relationship between quantum entropy inequalities and algorithmic quantum speedups. Our goal is to study the evolution of select information-theoretic quantities as a quantum state is evolved through a quantum circuit which implements a quantum algorithm that gains a speedup over classical counterparts. The motivation of this analysis is that it must be the fundamentally quantum nature of the state that fuels the speedup, and that the saturation, or degree of failure of entropic inequalities is a potential metric to gauge this fuel. We consider these quantities to be representative of how `quantum' the subsystem is behaving.

It is known that stabilizer states are efficiently simulable \cite{Gottesman:1998hu,Aaronson_2004}, and Clifford circuits can be classically simulated in polynomial (quadratic) time using the tableau representation. Hence any qubit\footnote{We limit our discussion in this paper to quantum computers consisting of qubits, i.e.\ with Hilbert spaces which are tensor products of $\mathbb{C}^2$. However, we note that there are obvious extensions of the concepts of stabilizer states and Clifford circuits to qudits, with base Hilbert space $\mathbb{C}^d$, $d>2$. Indeed, it is known that the Gottesman-Knill theorem extends to both Clifford circuits in qudit systems with $d$ equal to an odd prime, and to Gaussian circuits in (bosonic) continuous-variable systems \cite{Mari:2012ypq}. Separately, free fermionic systems, which can be mapped to ``matchgate'' qubit circuits, are also known to be efficiently classically simulable \cite{10.1145/380752.380785,Terhal:2002aii}.} algorithm with a quantum speedup must incorporate non-Clifford gates, either by preparing a non-stabilizer initial state or using gates such as the Toffoli or T gate in a circuit implementation. Because non-stabilizer states can be used to speed up computation, they are sometimes referred to as magic states \cite{Knill:2004ctr,Bravyi:2004isx}. On the other hand, it is known that entropies of stabilizer states satisfy Ingleton's inequality \cite{ingleton1971representation,Linden:2013kal,Walter_2014}. There are many non-stabilizer states that themselves have entropies which similarly obey Ingleton (for example, any state reachable from a stabilizer state by acting with local unitaries), but it is still natural to wonder whether violation of Ingleton can be used to diagnose quantum speedups.

Another subclass of quantum states, so-called ``holographic" states, have entropy vectors obeying monogamy of mutual information \cite{Hayden_2013,Bao_2015}. Holographic states are a strict subset of stabilizer states: as we discuss in the next section, states with large amounts of tripartite entanglement, such as the GHZ states \cite{Greenberger:1989tfe}, violate MMI.

We analyze three well-known algorithms (or subroutines of algorithms): Grover's search \cite{Grover:1996rk}, the quantum Fourier transform \cite{Coppersmith:2002skh}, and quantum phase estimation \cite{Kitaev:1995qy}. At every timestep in the algorithm, the state of the quantum computer is represented by its density matrix, from which can be derived all the density matrices of its subsystems, and hence their entanglement entropies, which we use to check inequalities. We find, perhaps as expected, that the saturation and failure of MMI and Ingleton's inequality do not suffice to diagnose speedup potential. None of the states evolved through circuits corresponding to our algorithms exhibit Ingleton-violated entropy vectors at any time, for example. Hence the evolution of the saturation of these inequalities does not give the full picture of how information is being exploited for these algorithms. Further analysis of saturation of different inequalities, or of more general information quantities, is required. 

The remainder of this paper is organized as follows. In Section \ref{sec:preliminaries}, we remind the reader of the definitions of entanglement entropy and the entropy vector, and write down the inequalities we will consider in the remainder of the paper. Section \ref{sec:method} briefly discusses our computational methods (our code is made available accompanying the paper). In Section \ref{sec:results}, we discuss three quantum algorithms and describe how entropic structure evolves as they are performed. Finally, we conclude and discuss future directions in \ref{sec:discussion}.

\section{Preliminaries}\label{sec:preliminaries}

\subsection{Quantum Structure}
A single qubit is defined as a two-state system completely described as a vector in Hilbert space $\mathcal{H}^1 = \mathbb{C}^2$. An error-free quantum register of $N$ qubits is then described by $\mathcal{H}=\mathbb{C}^2\otimes \mathbb{C}^2\otimes ... = \mathbb{C}^{2^N}$. Therefore, at any given point, the state of the quantum register as a whole is a single vector living in this Hilbert space $\ket{\psi}\in\mathcal{H}$, which, in general, is not a product state: $\ket{\psi}\ne\ket{\psi_1}\otimes\ket{\psi_2}\otimes ....$. One might also use the density matrix at a given point to describe the state, which is necessary for mixed states. This operator is defined as $\hat \rho=\ket{\psi}\bra{\psi}$. For a `perfect' quantum register described above, the entire state is, by definition, pure, but subsystems within the state, i.e., collections of $M<N$ qubits, are not necessarily pure. We need a way to describe the state of these subsystems, so an operator on the density matrix, the partial-trace operator, is defined such that it maps the density matrix describing the whole state to a density matrix describing a subsystem. Notationally, the partial trace is written as $\text{Tr}_B(\rho_{AB})$ if one is `tracing out' the $B$ subsystem of a larger system $\Hil_{AB}=\Hil_A\otimes\Hil_B$. Thus $\text{Tr}_B(\rho)$ is the reduced density matrix representing the state of the $A$ subsystem: 
\begin{equation}
    \text{Tr}_B(\rho_{AB})\equiv \rho_A = \sum_j(I_A\otimes\bra{j}_B)\rho_{AB}(I_A\otimes\ket{j}_B).
\end{equation}

Note that $I_A$ is on the left because $\rho_{AB}=\rho_A\otimes \rho_B$, we want to trace over $B$, and order matters. To trace out arbitrary specified qubits using this formula one would need to reorder them to the end, or adapt this formula to extend to $N$-partite systems by summing over only those that should be traced out. (We have implemented such a formula in our code, which we discuss in the next Section.) It is necessary, here, to trace out arbitrary qubits by their indices.

Given a density matrix, the Von Neumann entropy is defined as $S=-\tr(\rho\log_2\rho)$, which we can see is equivalent to the Shannon entropy (measured in bits) of the diagonal matrix elements when $\rho$ is written in the basis of its eigenvectors: $\rho = \sum_i p_i\ket{i}\bra{i}$. Computationally, we can find the entropy by either using some implementation of the matrix logarithm or by first finding the eigenvalues of the density matrix. We use the latter approach, so we can explicitly truncate numerical errors out of the computation.

Entropies of pure states are always zero, but the entropies of subsystems of pure states can be nontrivial, and indicate the entanglement structure of those subsystems with the rest of the system. The entropy vector is defined as $\vec{S} = \{ S(\operatorname{Tr}_{\bar{A}}) \} \, \forall \, A$, where $A$ is every possible subset of factors of the total system. With the tools to calculate the entropy vector, it is possible to analyze the sharing of information between subsystems, and between partitions of subsystems. In general, if order matters, a convention for the ordering of the entropy vector is required. Here, however, we treat it as a abstract map from the indices of the bits to the von Neumann entropies of their associated reduced density matrices. In the following we will not use the (ordered) entropy vector directly but just the collection of subsystem entropies it contains.
\subsection{Information Inequalities}\label{sub:ineqs}
We restrict our focus to only a few information-theoretic inequalities for this analysis. For every choice of subsystem, we can check entropic constraints on partitions of that subsystem. Consider the subadditivity inequality: 
\begin{equation}
    S(\rho_A)+S(\rho_B)-S(\rho_{AB})\geq 0.\label{eq:SA}
\end{equation} 
For a subsystem $\rho_{AB}$, any bipartition into $\rho_A$ and $\rho_B$ should obey Eq.\ \eqref{eq:SA}. We already know that subadditivity holds for all quantum states \cite{Araki:1970ba}, but we can now explicitly determine the saturation of subadditivity (SA) (i.e., the amount by which the left hand side exceeds the right hand side) for every partition of every subsystem. In principle this can be done for any quantum state, but we are interested, in particular, in how these constraints are saturated throughout the running of quantum algorithms which gain a speedup. We also check the tripartite inequalities of strong subadditivity (SSA) and monogamy of mutual information (MMI):
\begin{equation}
    S(\rho_{AB})+S(\rho_{BC})-S(\rho_{ABC})-S(\rho_B) \geq 0,\label{eq:SSA}
\end{equation}
\begin{equation}
    -I_3(A:B:C)\equiv S(\rho_{AB})+S(\rho_{BC})+S(\rho_{AC})-S(\rho_A)-S(\rho_B)-S(\rho_C)-S(\rho_{ABC}) \geq 0.\label{eq:MMI}
\end{equation}
$I_3$ is known as the \emph{tripartite information} \cite{Casini:2008wt,79902,1057469}. Note that these inequalities must be checked for all tripartitions of the subsystem. Strong subadditivity has been proven to hold \cite{Lieb_1973} for quantum states, but it is interesting to consider how close a particular state comes to saturating it. 

On the other hand, Eq.\ \eqref{eq:MMI} does not hold for arbitrary states, but does for ``holographic'' states \cite{Hayden_2013}. Some intuition may be gained by observing that the inequality \eqref{eq:MMI} is the sum of \eqref{eq:SSA} and $S(\rho_{AC})-S(\rho_A)-S(\rho_C)\ge 0$. However, this second inequality is guaranteed to be violated, because it is the negative of an instance of subadditivity \eqref{eq:SA}. Hence states which obey MMI must be far enough away from saturating SSA that there is still ``room'' to make up for the violation of the second inequality. The states which violate MMI are those with large amounts of tripartite information, so that SSA is nearly saturated. For example, consider the GHZ state on four qubits:
\begin{equation}
    \ket{GHZ_4}\equiv\frac{1}{\sqrt{2}}(\ket{0000}+\ket{1111}).
\end{equation}
Any tripartite subsystem of this state will fail the MMI inequality. Since MMI can fail for arbitrary states, we also keep track of the average failure saturation, and the ratio of failures to successes as metrics. 

Finally, for partitions of four pieces of each subsystem, we check Ingleton's inequality \cite{ingleton1971representation}, which is known to hold for all stabilizer states but not all quantum states:
\begin{equation}
    I(A:B|C)+I(A:B|D)+I(C:D)-I(A:B)\geq 0.
\end{equation}
Here, $I(X:Y)$ is the mutual information, and $I(X:Y|Z)$ is the conditional mutual information. Note that checking whether an entropy vector obeys an inequality entails checking all sets of entropies of subsystems, i.e.\ all permutations of $A,B,C,D$ ($D$ for Ingleton's only).

\section{Method}\label{sec:method}
We classically simulate a quantum computer by keeping track of the density matrix as it evolves. Running an algorithm on the simulated computer is effectively applying unitaries to the density matrix. The analysis is performed after every unitary is applied, constructing time-series data of the quantum state. At every time step, there are a number of checks of each inequality proportional to the number of partitions associated with it. Every subsystem is looked at, and there are a number of subsystems equal to the number of bipartitions of the qubits. Since we are interested only in the entropic dynamics of an algorithm itself rather than a particular gate set, we will be using the gate set which is pictorially given in the circuit diagram for each algorithm. Thus, each algorithm has a unique gateset for the purposes of this analysis. Further investigation with particular gate sets for a given computer will presumably give more resolution to these dynamics which would be more indicative of how computers with that gate set will behave. Algorithms which decompose circuits into specified gate sets (such as Clifford + $T$) would be required for such an analysis. The code for this entropy tracking and simulation can be found on Github here: \url{https://github.com/DylanJVA/QI_Research/}.

In the remainder of the paper, we will present plots which show the evolution of entropic quantities. If we think of a quantum circuit as a linear succession of gates, it might seem to make sense to plot these quantities after each successive gate is applied. However, in practice quantum circuits are drawn in layers, with the idea that if multiple gates happen in a layer they occur simultaneously. Our approach here is to focus on only the evolutions that can alter the entropic structure of a state. The x axis on our plots of saturation should therefore be read as circuit depth, \emph{not} number of gate applications. To recall that single-qubit gates do not change entanglement structure, consider the Schmidt decomposition of a bipartite state:\begin{equation}
    \ket{\psi}_{AB}=\sum_i \lambda_i\ket{u_i}\ket{v_i},
\end{equation} 
where the $\lambda_i$ are the strictly positive Schmidt coefficients, and $\ket{u_i}$ and $\ket{v_i}$ are orthogonal basis vectors that span the total Hilbert space. Two subsystems are entangled in the total state $\ket{\psi}_{AB}$ if and only if there is more than one $\lambda$. When applying a local gate to, say, system A, this operation can be represented as a local unitary $U\otimes I$ acting on the state:\begin{equation}
    (U\otimes I)\ket{\psi}_{AB}=\sum_i\lambda_i(U\ket{u_i})\ket{v_i}.
\end{equation}
Since $U$ only changes the basis of $\ket{u_i}$ without altering the Schmidt coefficients, the entanglement remains unchanged. 

In fact, if the dimension of $\ket{u_i}$ was larger than one qubit, meaning $U$ is a multi-qubit gate, there would still be no change in entanglement between the two partitions. However, since we are considering all $k$-partitions of $N$ qubits, only single-qubit gates won't change the full entropy vector. By this argument, any series of gates of the form $U_1 \cdot U_2 \cdot U_3...$, where each $U_i$ is a gate acting on a single qubit with identity on the rest will not change entanglement in the entropy vector. We condense every such consecutive series of gates into a single gate application for the purpose of this analysis. Therefore only controlled gates and gates which can only be applied to multiple qubits can not be condensed in a similar fashion, and these are the only gates for which entanglement tracking is nontrivial.
\section{Results}\label{sec:results}

\subsection{Grover's Algorithm}

\begin{figure}[H]
\centering
\begin{tikzpicture}
    \node[inner sep=0pt] (circuit) {
    \begin{minipage}{\textwidth}
    \[
\Qcircuit @C=1.8em @R=0.8em {
    \lstick{|0\rangle} & \gate{H} & \ctrl{1} & \gate{H} & \ctrlo{1} & \gate{H} & \qw & \cdots \ \ \ \ \ & \qw & \meter{} \\
    \lstick{|0\rangle} & \gate{H} & \ctrlo{1} & \gate{H} & \ctrlo{1} & \gate{H} & \qw & \cdots \ \ \ \ \ & \qw & \meter{} \\
    \lstick{|0\rangle} & \gate{H} & \ctrl{0} & \gate{H} & \ctrlo{0} & \gate{H} & \qw & \cdots \ \ \ \ \ & \qw & \meter{} \\
    & \vdots & \vdots & \vdots & \vdots & \vdots & & & & \vdots  \\ \\
    \lstick{|0\rangle} & \gate{H} & \ctrlo{1} & \gate{H} & \ctrlo{1} & \gate{H} & \qw & \cdots \ \ \ \ \ & \qw & \meter{} \\
    \lstick{|0\rangle} & \gate{X} & \gate{X} & \qw & \gate{Z} &\qw & \qw & \cdots \ \ \ \ \ & \qw \\ \\ \\ 
}
\]
    \end{minipage}
    };
    \draw[dashed] (circuit.north west) ++(4.5, -0.2) rectangle ++(0.5, -4);
    \node at ($(circuit.north west) + (3.5, +.3)$) [anchor=west] {goal bitstring};
    \node at ($(circuit.south) + (-0.6, +0.5)$) [anchor=north] {$\underbrace{\hspace{12em}}_\text{Grover iteration}$};
\end{tikzpicture}\caption{Circuit diagram for Grover's algorithm.\label{fig:grover}}
\end{figure}
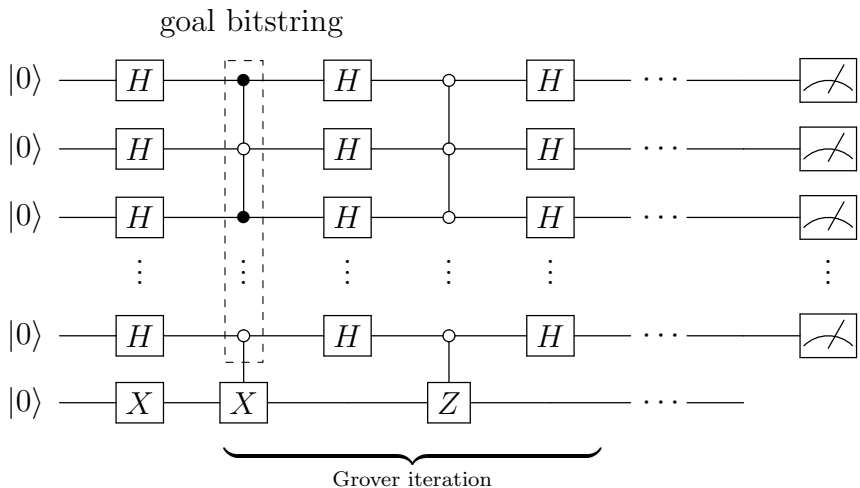

The first algorithm we consider is Grover's search \cite{Grover:1996rk}. Classically the optimal search of an $K-$ element unordered list does no better than a linear search with complexity $\mathcal{O}(K)$, checking every element until the goal is found. Grover's algorithm is known to have gate complexity $\mathcal{O}(\sqrt{K})$, which is an obvious speedup. This is a fundamental example because the classical method can intuitively be seen as optimal, and so the speedup must be quantum in nature. We know that the state begins as a classical one $\ket{00...}$ and ends in one as well--the computational basis state corresponding to the goal bitstring, when the alogorithm is successful--but enters the quantum regime somewhere in the middle. The implementation of the algorithm is characterized by the circuit diagram in Figure \ref{fig:grover}.

 For Grover's, the gate set is:
\begin{equation*}
H=\frac{1}{\sqrt{2}}
\begin{bmatrix}
1 & 1 \\
1 & -1
\end{bmatrix},\quad X=
\begin{bmatrix}
0 & 1 \\
1 & 0
\end{bmatrix},\quad Z=
\begin{bmatrix}
1 & 0 \\
0 & -1
\end{bmatrix},\end{equation*}
\begin{equation*}
C_\psi X=\ket{\psi}\bra{\psi}\otimes X + \sum_{j\neq \psi}\ket{j}\bra{j}\otimes I_{2^{n-1}}, 
\end{equation*}

\begin{equation*}
C_{00\ldots 0}Z=\ket{00\ldots 0}\bra{00\ldots 0}\otimes Z + \sum_{j\neq 0\ldots 0}\ket{j}\bra{j}\otimes I_{2^{n-1}},
\end{equation*}
where the summation is over all $K=2^{n-1}$ length-($n-1$) classical bitstrings, the two controlled operators are controlled on the first $n-1$ qubits and targeted on the last qubit, and $\psi$ represents the classical bitstring we are trying to find. We take these multiple-controlled gates as being part of the gate set, but in an actual implementation they would need to be decomposed into the gate set associated with the computer.

 We first consider the evolution of the saturation of strong subadditivity. Since strong subadditivity has been proven for all quantum states \cite{Lieb_1973}, we know it should never fail. For all results, we ran Grover's algorithm on $5$, $6$, and $7$ qubits, using $16$ Grover iterations, and amplifying the states $1101,11010,110101$ respectively. Note that $16$ is not the optimal number of iterations, but is chosen here to observe periodic evolution. In general, the number of iterations should be chosen to be the minimum such that the probability of measuring the goal state is highest, which is roughly $\approx \lceil \frac{\pi}{4}\sqrt{K}\rceil$ for $\log_2 K$ qubits in the goal state. It is important to note that of the algorithms presented, only Grover's allows room to change the number of iterations. The rest of the algorithms have gate depth which is determined exactly by the algorithm itself.
\begin{figure}[H]  
    \centering
    \includegraphics[width=0.8\textwidth]{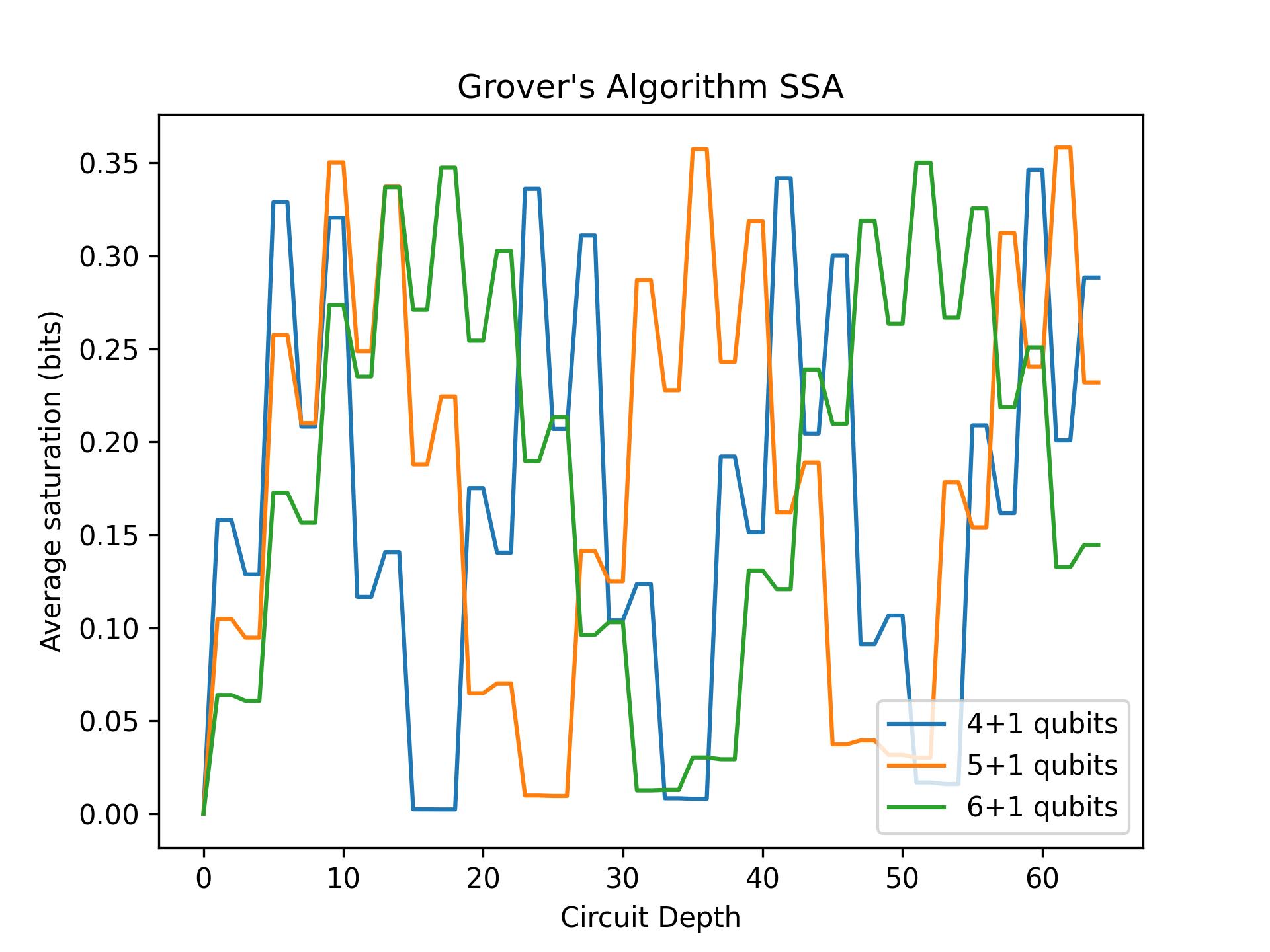}
    \caption{Strong subadditivity saturation during Grover's algorithm.}
\end{figure}

As expected, subadditivity was confirmed for all states during Grover's, and the saturation can be seen as periodic which reflects the nature of the algorithm. More gates are required with more qubits, and thus the periodicity of the evolution also grows. One can note that the saturation is at a minimum when the state is closest to the goal state.

Now we consider the monogamy of mutual information:
\begin{figure}[H]
    \centering
    \begin{minipage}{0.45\textwidth}
        \centering
        \includegraphics[width=\textwidth]{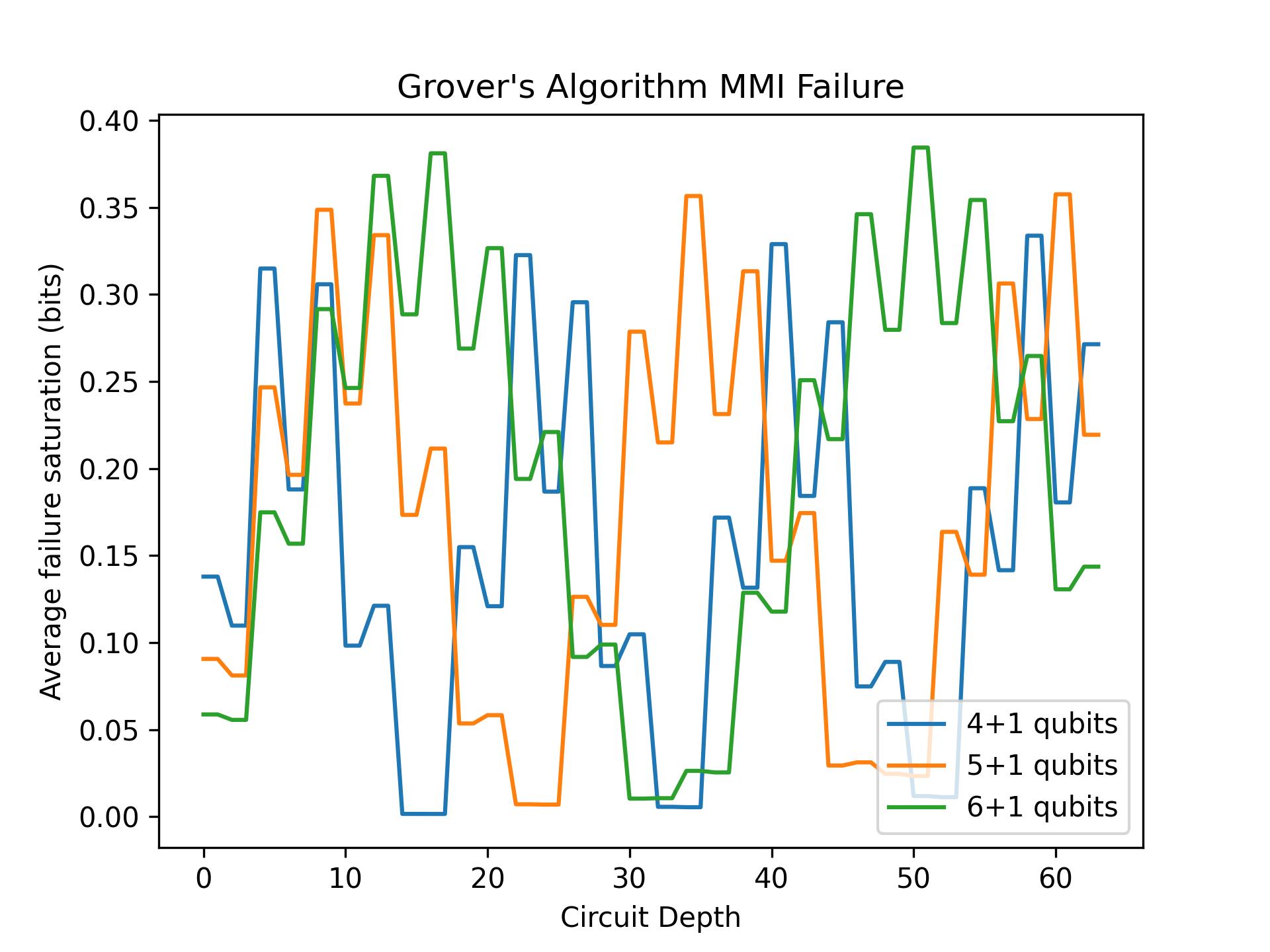} 
        \caption{The average (negative) saturation of subsystems which failed MMI during Grover's}
    \end{minipage}\hfill
    \begin{minipage}{0.45\textwidth}
        \centering
        \includegraphics[width=\textwidth]{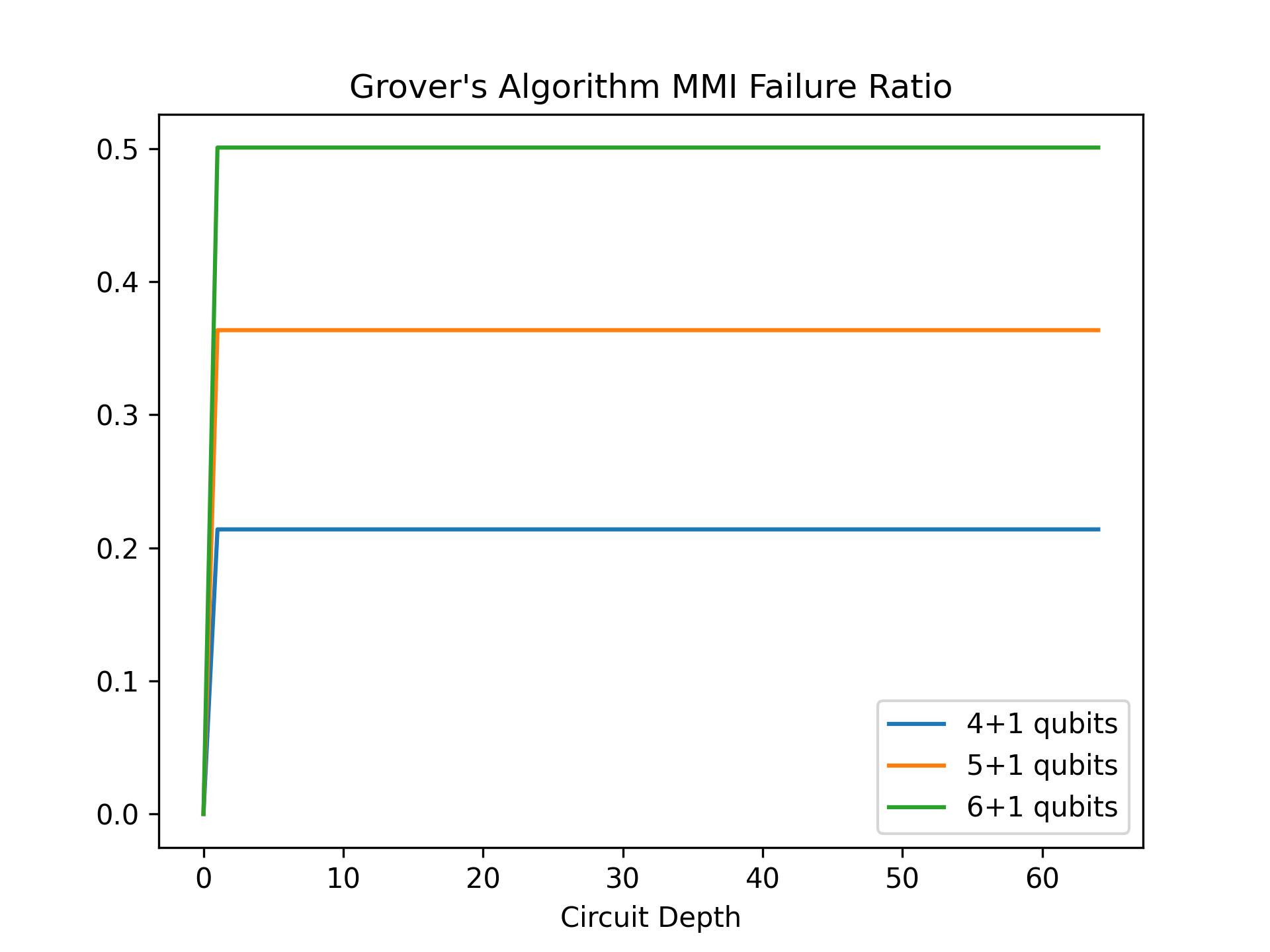} 
        \caption{The ratio of subsystems which failed MMI during Grover's. The ratios for 5,6, and 7 bits fixed at $0.214,0.364,$ and $0.501$ respectively}
    \end{minipage}
\end{figure}

We find that immediately after the first set of gates, states fail MMI, and that the ratio of failure is proportional to the number of qubits. It is clear that the MMI inequality for Grover's is either fully saturated (up to numerical precision), or fails completely, with periodic magnitude. This suggests that MMI is not an informative metric for Grover's as the algorithm introduces this failure nearly immediately, and the periodic evolution of the magnitude of this failure is related to the algorithm itself.

Turning to Ingleton's inequality:
\begin{figure}[H]  
    \centering
    \includegraphics[width=0.8\textwidth]{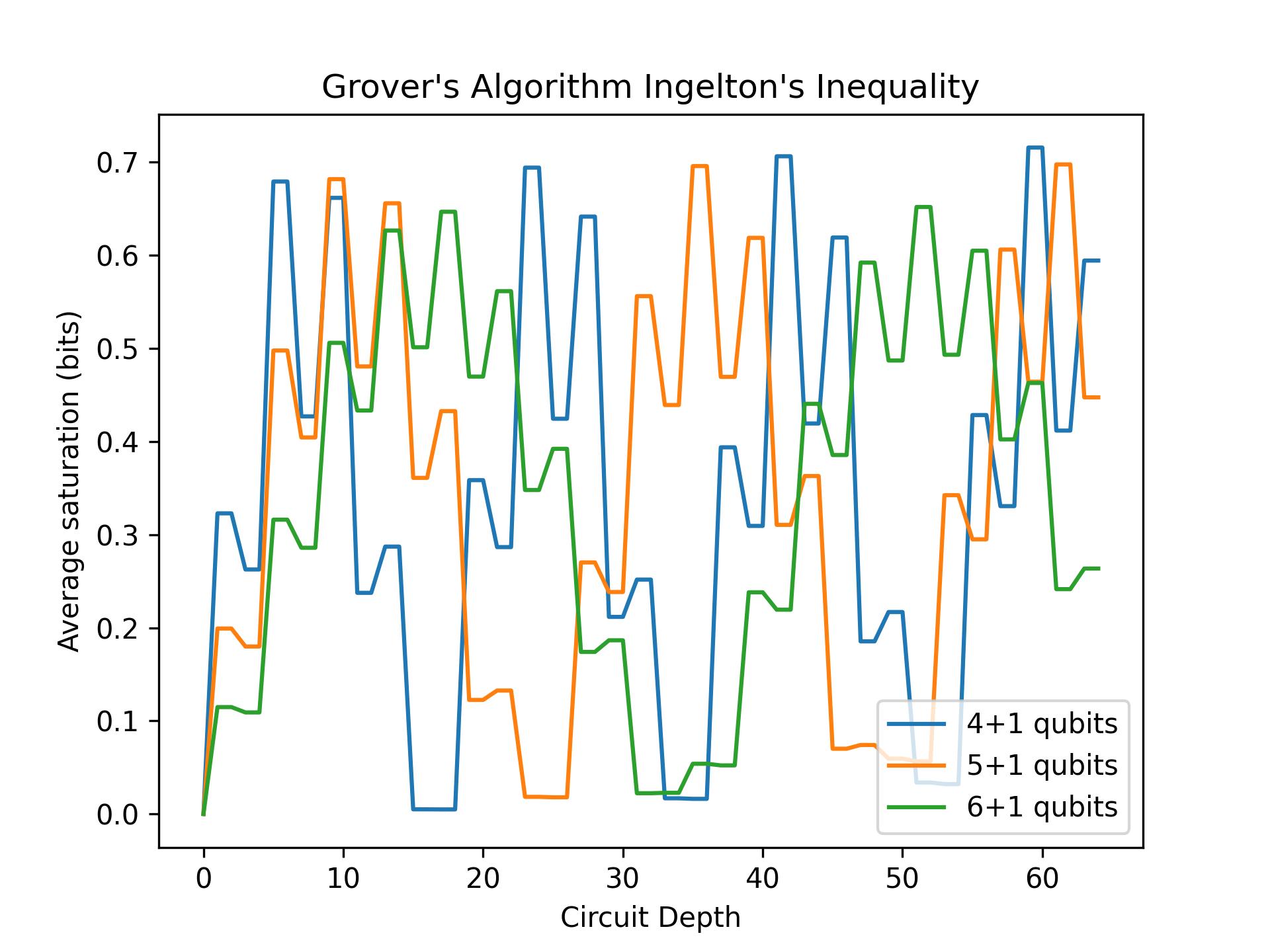}
    \caption{Ingleton's inequality saturation during Grover's algorithm.}
\end{figure}
In similar fashion to strong subadditivity, we see that no states were reached where subsystems failed Ingleton's inequality. Therefore we conclude that despite being obeyed by classically simulable states, it does not diagnose speedups effectively. 
\subsection{Quantum Fourier Transform}
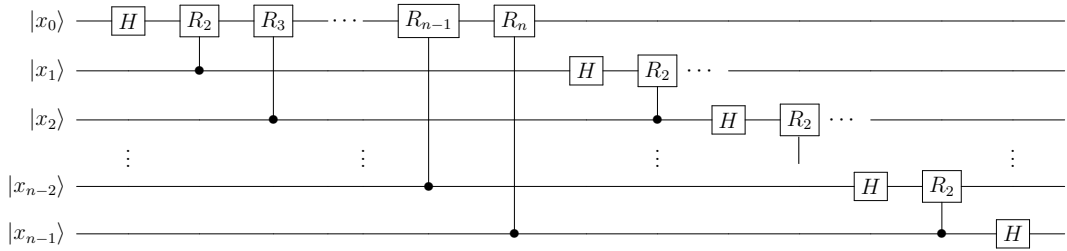
\begin{figure}[h]
\centering
\begin{tikzpicture}
    \node at (0,0) {
        \scalebox{0.7}{
            \Qcircuit @C=1.6em @R=0.8em {
                \lstick{|x_0\rangle} & \gate{H} & \gate{R_2} & \gate{R_3} & \qw &  \cdots \ \ \ \ \ & \gate{R_{n-1}} & \gate{R_n} & \qw & \qw & \qw & \qw &\qw & \qw &\qw &\qw \\
                \lstick{|x_1\rangle} & \qw & \ctrl{-1} & \qw & \qw & \qw  & \qw & \qw & \gate{H} & \gate{R_2} & \cdots \ \ \ \ \ \ \ \ & \qw & \qw & \qw & \qw & \qw\\
                \lstick{|x_2\rangle} & \qw & \qw & \ctrl{-2} & \qw & \qw & \qw &\qw & \qw & \ctrl{-1} & \gate{H} & \gate{R_2} &  \cdots \ \ \ \ \ \ \ \ & \qw &\qw & \qw\\
                & \vdots  & & & &\vdots & & & & \vdots & & & & & \vdots\\
                \lstick{|x_{n-2}\rangle} & \qw & \qw & \qw & \qw & \qw & \ctrl{-4} & \qw  & \qw & \qw & \qw & \qw & \gate{H} & \gate{R_2} &  \qw &\qw\\
                \lstick{|x_{n-1}\rangle} & \qw & \qw & \qw & \qw & \qw & \qw & \ctrl{-5}  & \qw & \qw & \qw & \qw & \qw & \ctrl{-1} & \gate{H} &  \qw \\
            }
        }
    };
    \draw[black, thin] (3.05, -0.14) -- (3.05, -0.5); 
\end{tikzpicture}
\caption{Circuit diagram for the quantum Fourier transform with a vertical line.\label{fig:qft}}
\end{figure}

While it does not necessarily gain a speedup over the FFT, the quantum Fourier transform (QFT) may be used in a subroutine by other algorithms, such as Shor's algorithm, which do gain a speedup. This implementation of the QFT uses $\mathcal{O}(4^N)$ gates which is equivalent to the classical gate complexity of the standard Discrete Fourier Transform. Note that the QFT takes as its input an arbitrary quantum state, which can itself fail some entropic inequalities. We present the circuit diagram in Figure \ref{fig:qft}.

Here we have a gate set:
\begin{equation*}
    H=\frac{1}{\sqrt{2}}
    \begin{bmatrix}
    1 & 1 \\
    1 & -1
    \end{bmatrix},\quad CR_k=
    \begin{bmatrix}
    1 & 0 & 0 & 0 \\
    0 & 1 & 0 & 0 \\
    0 & 0 & 1 & 0 \\
    0 & 0 & 0 & e^{i2\pi/2^k}
    \end{bmatrix},
\end{equation*}
where we assume we have access to every controlled $R_k$ gate.

Since this algorithm exactly fixes the number of gates for the desired output, the circuit depth will grow with input size. It is obvious that at each point in the algorithm, the state is highly dependent on the input. Thus, the evolution of these entropy relationships is also dependent on the input, and can be seen when applying the QFT on two random states. Only with some inputs do states fail MMI, but in no random input we tried did any fail Ingleton's\footnote{We could have tried a quantum state which is known to violate Ingleton, of course. However, empirically, nearly all random chosen quantum states have entropy vectors which obey Ingleton, and this will be true with higher and higher probability as qubit number increases.}. Below are the saturations of MMI with two different inputs. 
\begin{figure}[H]
    \centering
    \includegraphics[width=0.75\textwidth]{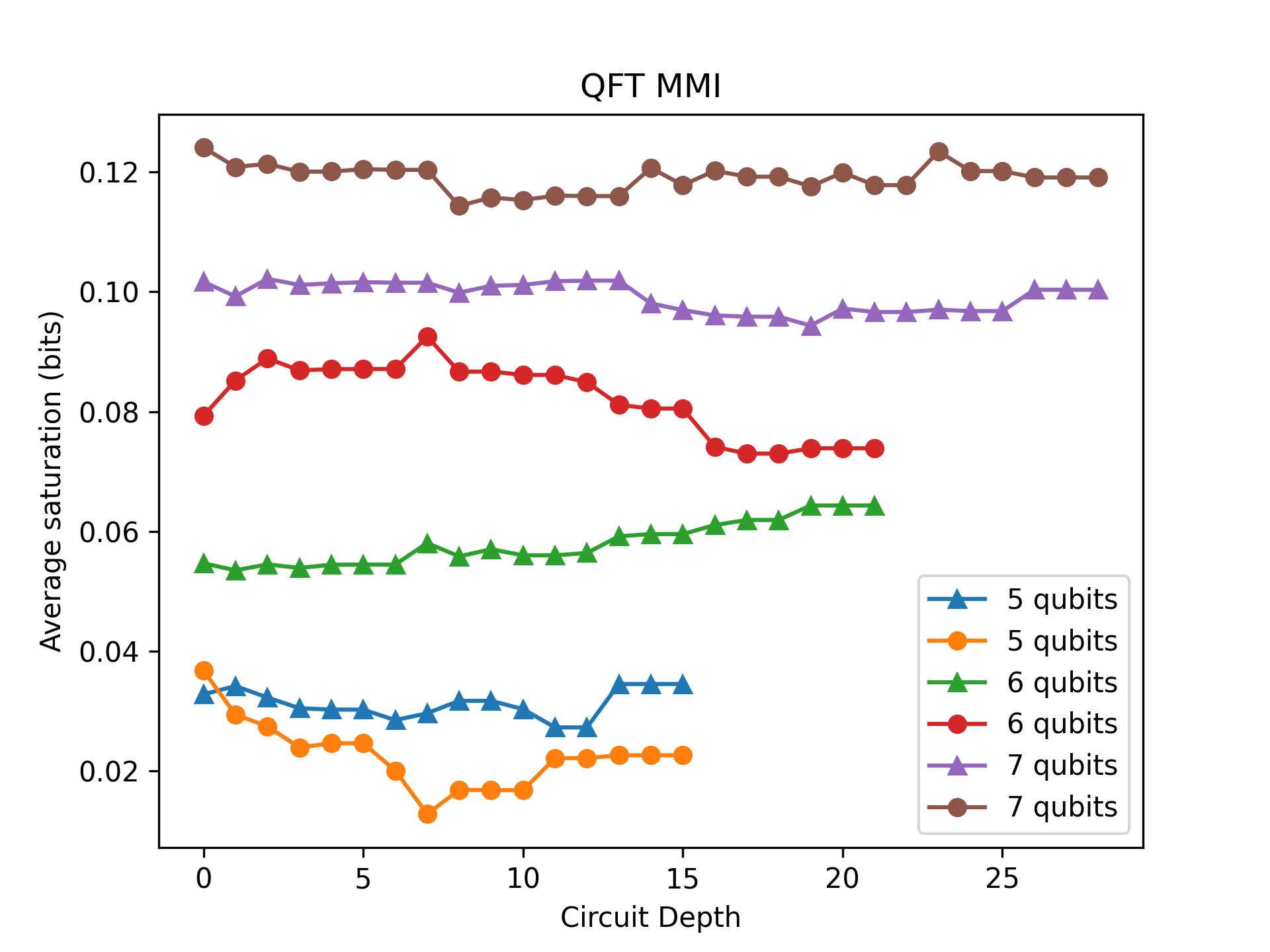} 
    \caption{Saturation of the monogamy of mutual information inequality during the QFT on two different random input vectors with varying sizes}
\end{figure}

Although the average MMI saturation is positive for the displayed inputs, some subsystems did fail the inequality. Only considering the failure saturation of subsystems that do not obey MMI also reflects largely varying results dependent on input. The failures can be compared below, using the same two states as before. In general with more qubits, there were more failing states, but this wasn't always the case; notice that the 5-qubit QFT had more failing states than the 6-qubit QFT for one particular input.  

\begin{figure}[H]
    \centering
    \includegraphics[width=0.75\textwidth]{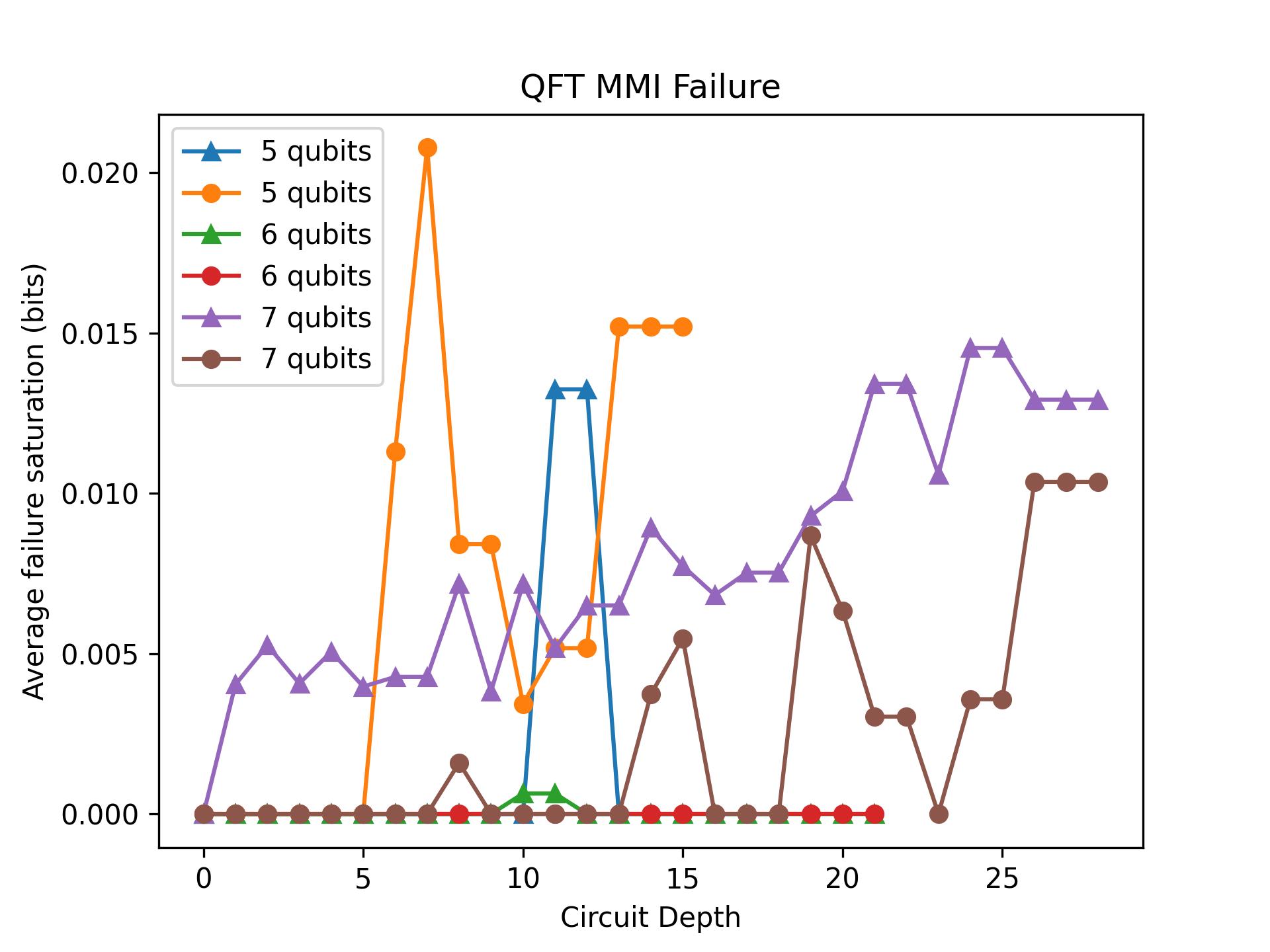} 
    \caption{The average (negative) saturation of only the subsystems which failed MMI during the Quantum Fourier transform on two different random input vectors with varying sizes}
\end{figure}

Ingleton's inequality had similar results as before, never failing. The evolution of the saturation also varies with input.

\begin{figure}[H]
    \centering
    \includegraphics[width=0.75\textwidth]{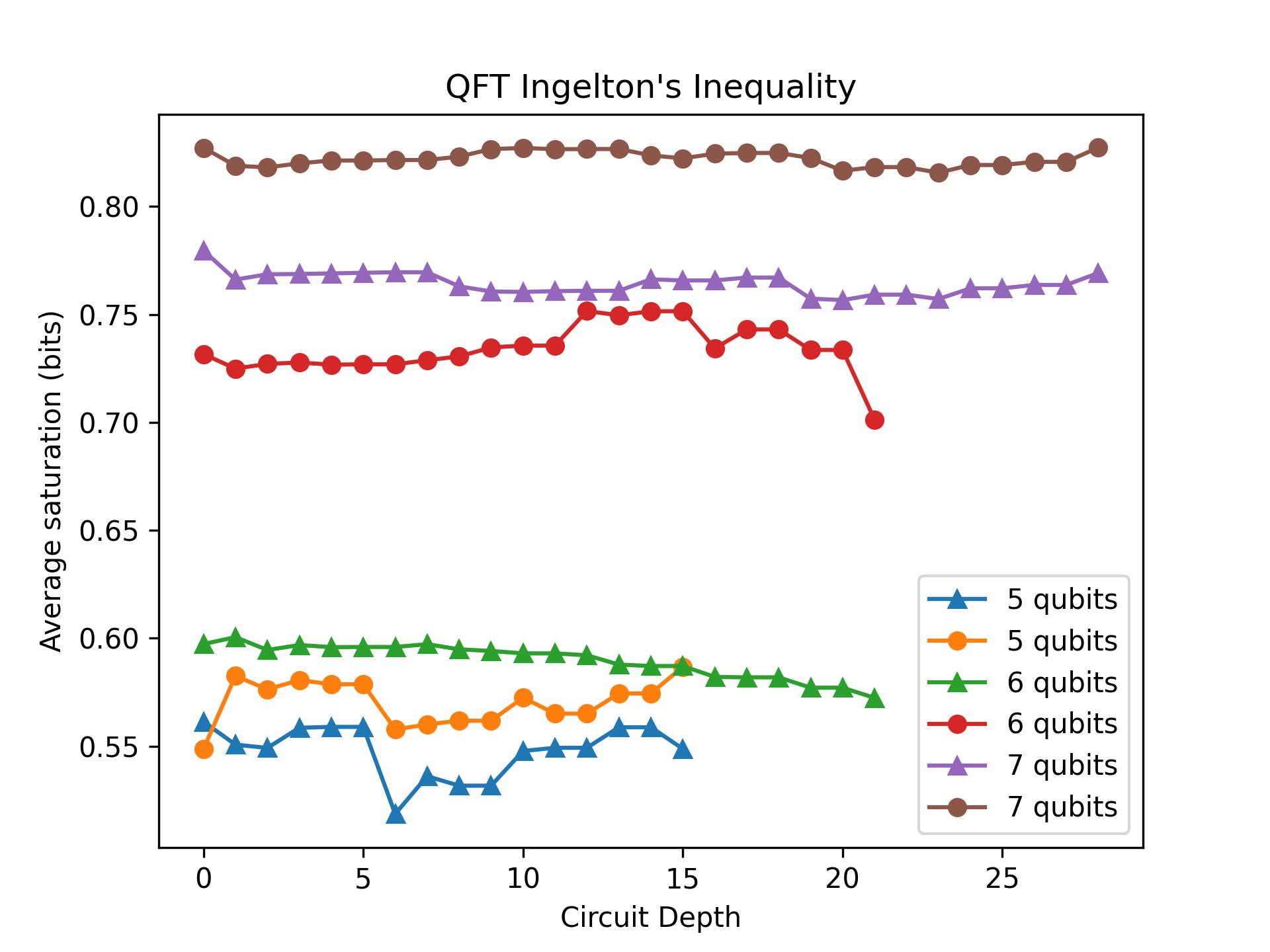} 
    \caption{Saturation of Ingleton's inequality during the QFT on two different random input vectors with varying sizes}
\end{figure}

\begin{figure}[h]
\centering
    \scalebox{0.8}{
\Qcircuit @C=1.6em @R=0.8em {
    \lstick{|0\rangle} & \gate{H} &\qw &\qw &\qw &\cdots \ \ \ \ \ &\ctrl{5}  &\multigate{4}{QFT^{-1}} & \qw &\ \ \ \ \ \ \ \ \ \ \ \ \ \ket{0}+e^{2\pi i 2^{t-1}\phi}\ket{1}\\
    & \vdots &  & & & & \\ \\
    \lstick{|0\rangle} & \gate{H} & \qw & \ctrl{2} &\qw &\cdots \ \ \ \ \ &\qw &\ghost{QFT^{-1}}&\qw &\ \ \ \ \ \ \ \ \ \ \ \ket{0}+e^{2\pi i 2^1\phi}\ket{1}\\
    \lstick{|0\rangle} & \gate{H} & \ctrl{1} & \qw &\qw & \cdots \ \ \ \ \ &\qw &\ghost{QFT^{-1}} &\qw &\ \ \ \ \ \ \ \ \ \ \ \ket{0}+e^{2\pi i 2^0\phi}\ket{1}\\
    \lstick{|\psi\rangle} & \qw & \gate{U^{2^0}} &  \gate{U^{2^1}} & \qw &\cdots \ \ \ \ \  & \gate{U^{2^{t-1}}}  &\qw &\qw&\ket{u}
    }}\caption{Circuit diagram for quantum phase estimation. \label{fig:qpe}}
\end{figure}
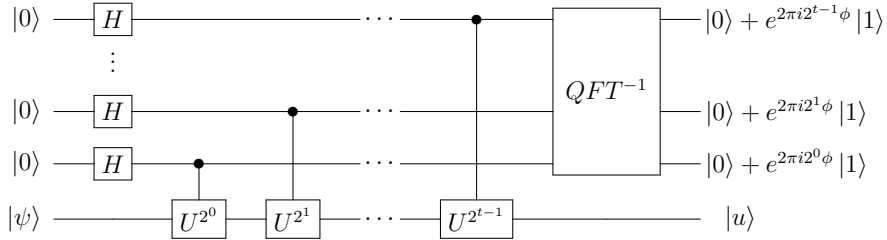

As mentioned, the QFT is used as a step in the quantum phase estimation algorithm, which may be used to find the phase of a particular unitary given one of its eigenvectors. Oracular access to the unitary is required, and thus controlled versions are treated as part of the gate set. The circuit is described by the diagram in Figure \ref{fig:qpe}.

One performs the quantum phase estimation algorithm on a state $\ket{\psi}$ which is an eigenvector of some unitary $U$ for which we want to find the phase, such that $U\ket{\psi}=e^{i\theta}\ket{\psi}$. The number of extra qubits necessary, $t$, defines the precision of the output.

 Phase estimation is an example of an algorithm which, other than the QFT subroutine, does not utilize or induce entanglement in its application. Recall that we know the unitary in general will apply a phase $e^{i\theta}$ to its eigenvector, and that it only does so controlled on the $\ket{1}$ state. To understand this, consider an arbitrary controlled $U$ on a state $\ket{\phi}$ controlled by a qubit in state $\alpha\ket{0}+\beta\ket{1}$:\begin{equation}
    \alpha\ket{0}\ket{\phi}+\beta\ket{1}e^{i\theta}\ket{\phi}=(\alpha\ket{0}+\beta e^{i\theta}\ket{1})\ket{\phi}.
\end{equation}
The phase can be moved through to a relative phase on the control qubit, and the target state is unchanged. Since this is a local transformation on just the control, by the arguments given above, this will not change entanglement. That is, if the controlled unitary acts on a product state $\ket{\psi}_{\rm control}\otimes\ket{\phi}_{\rm target}$, where the target state $\ket{\phi}$ is an eigenvector of the unitary, no entanglement will be created. Consequently, no entanglement is induced between the auxiliary qubits of the QPE and those which the unitary acts upon, even with successive applications of $U$ ($U^2, U^4, ...$). We expect to see no change in the saturation of these inequalities for algorithms that preserve entropies and entanglement structure. This was the case our results, where, for example, the MMI inequality saturation only changed once the inverse QFT is initiated. 

\begin{figure}[H]
    \centering
    \begin{minipage}{0.45\textwidth}
        \centering
        \includegraphics[width=\textwidth]{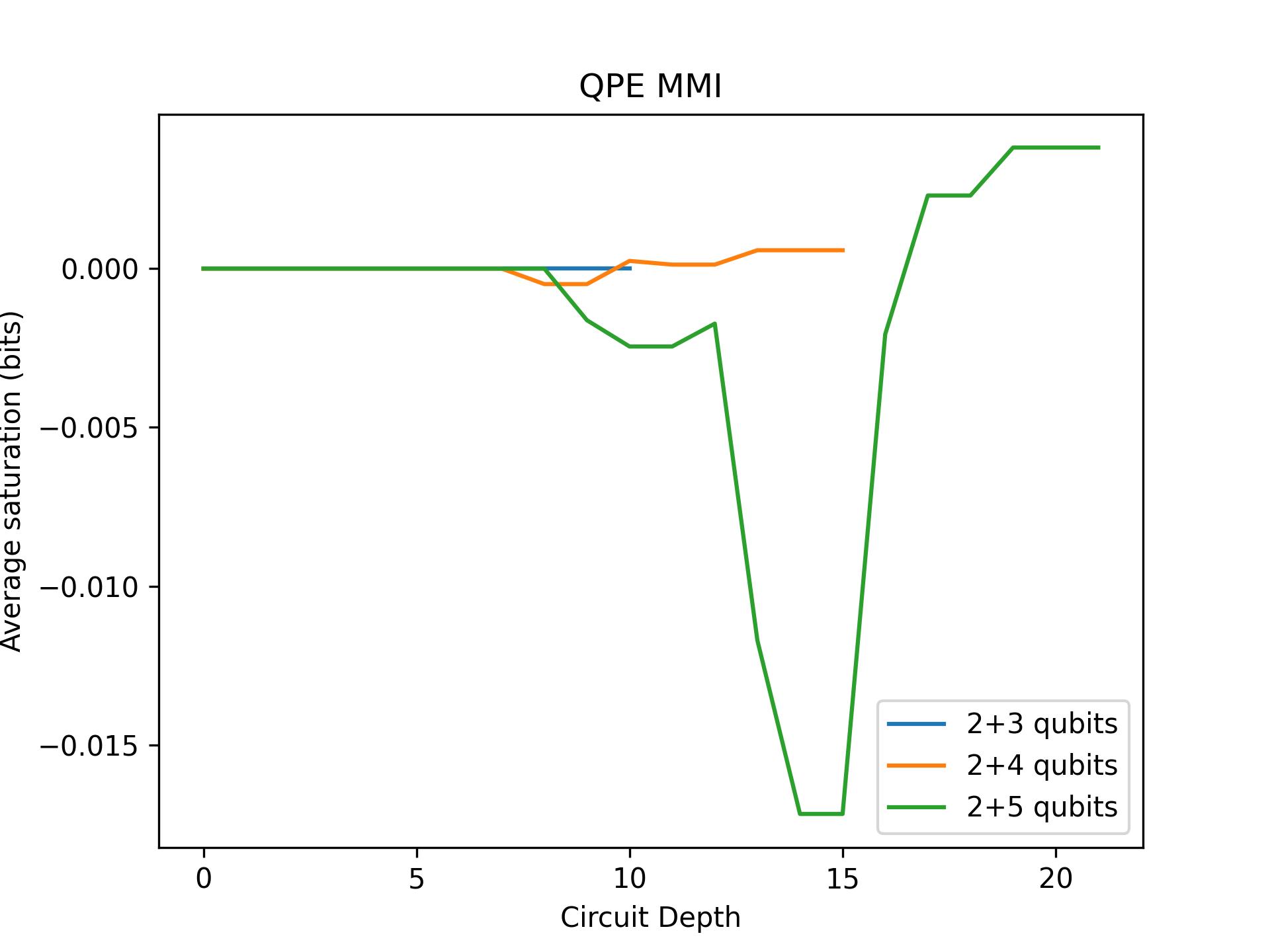} 
    \end{minipage}\hfill
    \begin{minipage}{0.45\textwidth}
        \centering
        \includegraphics[width=\textwidth]{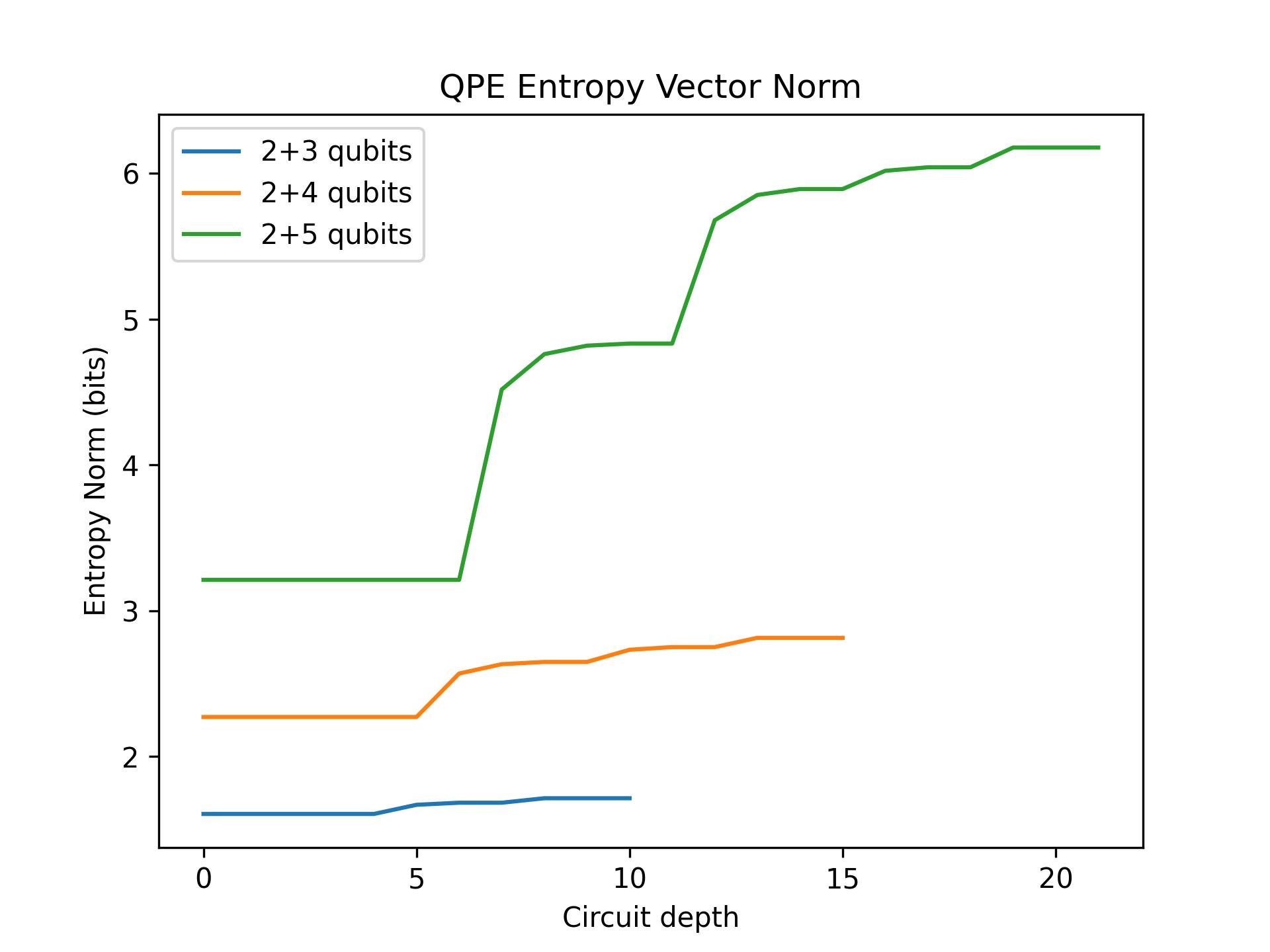} 
    \end{minipage}
    \caption{Saturation of monogamy of mutual information (left) and entropy vector norm (right) during quantum phase estimation. No entanglement is induced until the inverse QFT begins. In the case of $3$ auxiliary qubits, there is no change in MMI despite entanglement being created.\label{fig:norm}}
\end{figure}

Recall that MMI is an inequality relating entropies of a $3$-party system. Since the auxiliary qubits are never entangled with the qubits constructing the eigenvector, the MMI inequality will not change from zero. We can see that the entanglement structure is nevertheless changing despite MMI saturation remaining constant by looking at the norm of the entropy vector, shown in the right panel of Figure \ref{fig:norm}. Note that the norm we use for the entropy vector is the standard $2$-norm. Since there are more partitions of all qubits which divide the two (entangled) qubits in $\ket{\psi}$, the norm will be larger for larger qubit number despite the same amount of fundamental entanglement. Thus, the overall value of the norm is of less interest than its evolution. 

\section{Discussion}\label{sec:discussion}

We first emphasize a negative conclusion: in no case did we find quantum speedup correlated with violation of Ingleton's inequality. Hence the entropic inequalities we checked were insufficient to diagnose the presence or absence of a quantum speedup. That is, we were unable to make precise our initial intuition behind Ingleton's inequality being able to `diagnose' the potential fuel for quantum speedups. 

It is true that our simulation of quantum algorithms did not descend all the way to the level of a full circuit using a specified universal gate set. It is certainly likely that there exist some universal gate sets in which Ingleton's is ubiquitously violated by typical gate applications; but we expect this violation to be ultimately uncorrelated with any quantum speedup, since different quantum platforms could use different gate sets but all universal ones should have (up to logarithmic corrections) the same performance on a given task \cite{kitaev1997quantum}.

Interestingly, we have seen that in some algorithms, monogamy of mutual information fails during the course of the circuit. MMI has most often been considered in the context of holographic quantum states \cite{Bao_2015}, but it is not clear that the circuits and algorithms we considered should have any particular relation to holography. In Section \ref{sub:ineqs} above, we emphasized a more straightforward interpretation of MMI violation as requiring positive tripartite information and hence near-saturation of SSA subinequalities. It would be interesting to pursue this interpretation further.

Further investigation of other information-theoretic quantities is warranted. It seems intuitive that quantum speedup should be independent of a local basis choice, and depend only on the entanglement between factors of the Hilbert space representing physical qubits. Can we find better information quantities which more accurately diagnose, for example, the presence or absence of magic \cite{Knill:2004ctr,Bravyi:2004isx}? Or which identify which states are in the same LOCC classes \cite{Dur:2000zz,Verstraete_2002}? Many other information-theoretic quantities have been proposed and could be investigated as a state evolves through a quantum circuit: for example, tangle \cite{Coffman:1999jd} and negativity \cite{Vidal:2002zz}.

More generally, we note that algorithmic speedup compared to classical implementations is not the only possible benefit to be gained by quantum devices. For example, quantum teleportation \cite{PhysRevLett.70.1895} can be achieved using only Clifford circuits (indeed, with only Pauli gates), measurement and classical communication. Entanglement still seems essential to the protocol, but it requires only the Bell-type entanglement which we know does not suffice for quantum speedups. It would be interesting to further characterize the nature of the needed entanglement using entropic quantities.

\acknowledgments

The authors thank ChunJun Cao, Cynthia Keeler, and William Munizzi for helpful discussions.

\bibliographystyle{JHEP}
\bibliography{paper}
\end{document}